\documentclass[onecolumn, tighten]{aastex631}
\usepackage{nicefrac}
\usepackage{microtype}
\usepackage{booktabs}
\usepackage{rotating}
\usepackage{xcolor}
\usepackage[utf8]{inputenc}
\usepackage{amsmath}

\DeclareUnicodeCharacter{2212}{\ensuremath{-}}

\submitjournal{Frontiers in Astronomy and Space Sciences}
\shorttitle{Planetary Observation in Radio band}
\shortauthors{Bagheri et al.}

\begin{document}

\title{Infrared-Radio-follow-up Observations for Detection of the Magnetic Radio Emission of Extra-Solar Planets: A New Window to Detect Exoplanets}

\author{Fatemeh Bagheri}
\affil{NASA Goddard Space Flight Center, Greenbelt, MD, USA}
\affil{Department of Physics, University of Texas at Arlington, Arlington, TX, USA}
\correspondingauthor{Fatemeh Bagheri}
\email{fatemeh.bagheri@nasa.gov}

\author{Ramon E. Lopez}
\affil{Department of Physics, University of Texas at Arlington, Arlington, TX, USA}

\author{Amir Shahmoradi}
\affil{Department of Physics, University of Texas at Arlington, Arlington, TX, USA}

\begin{abstract}

%%% Leave the Abstract empty if your article does not require one, please see the Summary Table for full details.

\noindent There are several methods for indirectly detecting exoplanets, such as transit, radial velocity, astrometry, and the conventional gravitational microlensing approach. These methods rely on observing the effects of exoplanets on the emission or motion of observed stars. All these techniques have focused on the optical or infrared domains.  However, an alternative method for exoplanet detection via microlensing events involves planets orbiting the source star, creating a binary source system. In this study, we explore a novel approach to detecting and studying exoplanets exclusively through their radio emissions resulting from magnetospheric processes. We propose utilizing the Roman telescope as a survey observer to detect microlensing events. Subsequently, we investigate the potential for detecting planetary radio signals through follow-up observations of these microlensing events in the radio band using the SKA telescope. This method is viable due to the comparable radio emission levels of exoplanets and their parent stars, unlike optical and infrared emissions. We conduct a Monte Carlo simulation to replicate the observations by the Nancy Roman Telescope, followed by a follow-up observation in radio frequencies using the SKA telescope. We determine that approximately 1317 exoplanets exhibit detectable signals by the SKA telescope during the 7-season observations by the Nancy Roman Telescope.
This result indicates that such a method cannot only facilitate the direct detection of exoplanets but also enable the measurement of their magnetic field strength through analysis of their radio emissions.

\end{abstract}

\section{Introduction}
\begin{figure}[ht!]
\centering
\includegraphics[width=65mm]{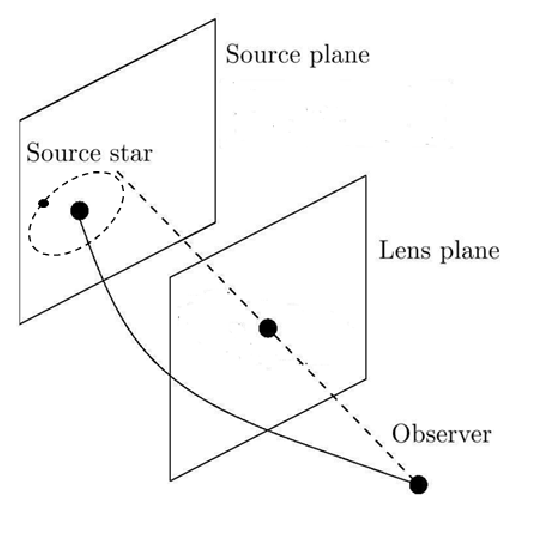}
\caption{A schematic view of a binary-source microlensing event.
\label{fig:Binsource}}
\end{figure}

\noindent A planetary magnetic field is one of the few observable quantities that contain information about the properties of the planetary interior since the internal dynamo produces it. Further, the extent to which exoplanets retain their atmospheres over geological time scales depends on whether they are exposed directly to stellar winds. Also, planetary magnetic fields could potentially serve as indicators of plate tectonics. Therefore, information about the magnetic fields of extrasolar planets is essential to assess their habitability. So far, our knowledge of planetary magnetic fields is limited to our solar system's planets, which are not representative of the vast array of possible planets, such as hot Jupiters \citep{lammer2009makes, lingam2021life, cockell2016habitability, kane2021planetary}. \\
\\
But how can we observe and measure the magnetic field of exoplanets?  The magnetospheres of Earth, Jupiter, Saturn, Uranus, and Neptune all show evidence of nonthermal continuum radiation emission resulting from the solar wind's interaction with the planet's magnetic field \citep{zarka1998auroral}. The electron cyclotron maser instability (ECMI) is a phenomenon that occurs in plasmas where electrons emit coherent electromagnetic radiation due to confinement by magnetic field lines along which they travel, gyrate, and accelerate \citep{melrose1982electron, dulk1985radio}. The ECMI is the most efficient radio emission mechanism, dominating all other emissions (beam-plasma instabilities and others). Based on our knowledge of the solar system planets, the average ECMI is related to the total solar wind power and the magnetic field of the planet by using the Radiometric Bode’s Law (RBL) model \citep{zarka2001magnetically, zarka2004fast, zarka2007plasma}. Therefore, observations of this emission type can indirectly measure and classify the planetary magnetic field. This quantification of the magnetic field gives us constraints on internal structure models and planetary rotation, informing us about the evolutionary history of the planet and host system \citep{hess2011model}. Furthermore, the exoplanets’ radio emission may reveal the presence of exomoons around them. For example, it has been shown that part of the Jovian decameter radio emission is “controlled” by the satellite Io \citep{bigg1964influence}. \citet{noyola2014detection} argues that the Io-Jupiter interaction could also be found in exoplanet-moon pairs, and the exoplanets’ radio emissions could be used to detect such systems directly.\\
\\
A multitude of exoplanet detection techniques rely on indirect optical/infrared wavelength observations \citep{wright2012frequency, knutson2014friends, mandel2002analytic, seager2002constraining, traub2010direct, guyon2005exoplanet, gaudi2012microlensing, bagheri2019detection}. However, measurements of the planetary magnetic fields require direct observations at the low-frequency radio wavelengths. Despite efforts to detect exoplanets in the radio band via ground-based radio telescopes, no promising detection of planetary radio emissions has been successful \citep{murphy2015limits, lazio2018magnetic, turner2021search}, primarily due to the low sensitivity of ground-based radio telescopes. Consequently, direct confirmation of the existence of exomoons has so far remained impossible.\\
\\
In this work, we combine two different exoplanet detection methods to make the radio emission of exoplanets detectable. The novel idea behind this study is that during a microlensing event, light intensity will be magnified at all frequencies. Therefore, a microlensing event can serve as a natural light magnifier. The idea is to use an optical/infrared telescope to detect a microlensing event that would trigger follow-up observations by radio telescopes upon detection. If the source star has at least one exoplanet in microlensing events, we can directly measure the light emitted from the exoplanet (Figure \ref{fig:Binsource}). The magnified light of the exoplanet could be observable in the infrared wavelength as we have shown in \citep{bagheri2019detection}. However, the radio band is the most desirable wavelength window to detect exoplanets \citep{vedantham2020coherent, viswanath2020statistical, foster2020corona} since the planetary radio emission, which includes the effects of exomoons, may dominate the signal from the parent star \citep{cauley2019magnetic, zarka2004fast}. This method will enable us to measure the magnetic fields of exoplanets and provide us with information about the existence of exomoons. To show this, we simulate the Nancy Grace Roman Space Telescope (previously named Wide Filed Infrared Space Telescope, WFIRST) as an infrared survey telescope to alert the Square Kilometre Array (SKA) radio telescope to the occurrence of microlensing events for follow-up observations. 

\section{Methodology}
\subsection{Microlensing events}
Microlensing, a powerful tool in the study of astrophysics, involves the gravitational bending of light by massive objects, typically stars, within the lensing galaxy. This phenomenon, predicted by Einstein's theory of general relativity, manifests as a temporary increase in the brightness of a background source, such as a distant star, as the foreground lensing object passes in front of it. The gravitational field of the foreground object acts as a lens, magnifying and distorting the light from the background source. In lensing, the magnification, denoted by $A$, represents the factor by which the flux of the source is amplified during the microlensing event. Therefore it is given by the ratio of the image size to the source size:
\begin{equation}
A = \frac{\text{Image size}}{\text{Source size}}~.
\end{equation} 
For point source magnification in microlensing, where the source size is negligible compared to the Einstein radius ($R_E$), the magnification can be simplified as:
\begin{equation}
    A = \frac{u^2 + 2}{u\sqrt{u^2 + 4}} 
\end{equation}
where $u$ is the impact parameter, defined as the closest approach of the source to the lens normalized by the Einstein radius \citep{gaudi2012microlensing}. % A magnification of The magnification is crucial for understanding the light curve of the microlensing event, which provides information about the lensing system, including the mass and distance of the lensing object. 
The lensing system may comprise either a single object (single-lens) or two objects (binary-lens), with the possibility of additional components \citep{danvek2019triple}. The presence of two lenses introduces additional complexity to the gravitational lensing phenomenon, leading to more intricate and often asymmetric light curves. Source systems can also consist of one or more objects. The conventional technique utilized in gravitational microlensing for exoplanet detection relies on the premise that the lens star hosts an exoplanet, constituting a binary lens system. Alternatively, another microlensing approach involves a scenario where the planet orbits the source star, resulting in a binary source configuration \citep{bagheri2019detection} as illustrated in Figure \ref{fig:Binsource}. In this study, we simulate binary-source microlensing events by assuming both single-lens and binary-lens configurations. We consider the finite source effect for the source star while neglecting limb darkening.  
\subsection{Source and Lens Stars Parameters}
\noindent To characterize the star mass density within the Milky Way galaxy, we rely on the B\'{e}sançon model, extensively discussed in previous works \citep{spergel2015wide, robin2003synthetic, gardner2014n}. This model presents the distribution of matter in our galaxy as a superposition of eight thin disk structures with varying ages, a thick disk component, and a central (old) bar structure consisting of two components. Our choice is the updated model proposed by \citep{gardner2014n}. We select the model parameters fitted to a two ellipsoid bar. This refined model, utilizing star distribution data from the Hipparcos catalog, has proven effective in interpreting microlensing data from the EROS collaboration in the direction of the spiral arms \citep{moniez2017understanding}. When generating binary lenses, the mass ratio between the two lenses is determined from the distribution function proposed by \citep{duquennoy1991multiplicity}. We choose the semi-major axis $s$ of the binary orbit of microlenses from the \"{O}pik's law where the distribution function for the primary- secondary distance is proportional to $\rho(s) = dN/ds \propto s^{-1}$ in the range of $[0.6, 30]$ au.

\subsection{Exoplanet Parameters and their Occurrence}

\noindent In our simulations, we categorize the characteristic parameters of exoplanets into two primary groups: rocky planets and Jovian planets. According to the conventional planet formation theory, rocky planets are situated within the snow line of their parent star \citep{kennedy2006planet, kennedy2008planet}. The snowline of a star, also known as the frost line or ice line, refers to the distance from the star where volatile compounds, such as water, can condense into solid form due to the decrease in temperature. Beyond the snowline, the temperature drops low enough for these volatile compounds to freeze, forming icy grains or particles \citep{kennedy2008planet}. This snowline can be expressed with the snowline of the Sun as
\begin{equation}
R_{\text{SL},\star} = R_{\text{SL}_\odot} \times \frac{M_{\star}}{M_{\odot}}~,
\end{equation}
where $M_\star$ is the mass of the parent star \citep{gould2010frequency}. For Jovian planets, as they can migrate towards their parent star from distant orbits (as discussed by \citep{murray1998migrating, papaloizou2005planet}, we set the maximum distance from the parent star at $10$ au. This range is consistent with observations from MOA microlensing studies, indicating at least one bound planet per star within this range, represented as $a = 0.01 - 10$ au \citep{microlensing2011unbound}. We assume that the distribution of exoplanet semi-major axes follows the \"{O}pik's.\\
\\
The planetary mass range in our simulation is $0.002 - 10 M_J$ \citep{schlaufman2018evidence}. For planets with masses $> 100~M_\oplus$ orbiting host stars with masses greater than $0.6$ solar masses, we utilized the correlation between giant planet occurrence and host star mass and metallicity as outlined by \citep{johnson2010giant, fulton2021california}. Following the occurrence dependencies discussed in \citep{fulton2021california, santerne2016sophie, santerne2016sophie, wittenmyer2020cool, winn2024planet}, we adopted their result of planet occurrence. We summarized the frequencies of various exoplanet types from our simulation in Table \ref{tab:fvsMp}.\\
\\
Generally, exoplanets exhibit a wide range of mass-radius relationships, reflecting their diverse compositions and evolutionary histories. However, a commonly used empirical relationship for estimating the radius of exoplanets based on their mass and incident flux is the following power-law scaling relation
\begin{equation*}
   \frac{R_p}{R_\oplus} = 1.78~ \Big(\frac{M_p}{M_\oplus}\Big)^{0.53}~\Big(\frac{F}{erg~ s^{−1}~cm^{−2}}\Big)^{−0.03}~~~~~~ \text{for} ~~M_p < 150~M_\oplus 
\end{equation*}
\begin{equation}
   \frac{R_p}{R_\oplus} = 2.45~ \Big(\frac{M_p}{M_\oplus}\Big)^{-0.039}~\Big(\frac{F}{erg~ s^{−1}~cm^{−2}}\Big)^{0.094}~~~~~~ \text{for} ~~M_p > 150~M_\oplus  
\end{equation}
where $F$ is the bolometric incident flux from the parent star and can be calculated by
\begin{equation}
    F = \sigma T_{\text{eff}}^{4} ~\frac{R^2_\star}{a^2}~ \sqrt{\frac{1}{1-e^2}} ~,
\end{equation}
where $R_\star$ is the stellar radius, $T_\text{eff}$ is the effective stellar temperature, $a$ is the semi-major axis, and $e$ is the orbital eccentricity \citep{weiss2013mass}.

\begin{longtable}{cccc}
\hline\hline
Class  & Planet Mass $[M_{J}]$	&	Orbital Period $[day]$ 		&Frequency $\%$ \\
\hline
Earth               & 0.002-0.007	&	0.8 - $10^{4}$	&	0.018	\\
Super Earth         & 0.007-0.018	&	0.8 -  $10^{4}$	&	0.042	\\
Small Neptune       & 0.018-0.033	&	0.8 $<$	      &	0.058	\\
Large Neptune       & 0.033-0.077	&	10 $<$	      &	0.167	\\
sub giant           & 0.077-0.3	    &	10 $<$	    &	0.010  \\
giant               & 0.3-5	        &	10 $<$	    &	0.040  \\
super giant         & 0.3-5	        &	10 $<$	    &	0.49  \\
Hot Jupiter         & 0.3-10	    &	0.8 - 10 	    &	0.07	\\
\hline
\caption{Summary of the planet occurrence frequencies in our simulation, as a function of planet mass and period.} \label{tab:fvsMp}
%\end{tabular}
\end{longtable}

\subsection{Estimation of Exoplanets Radio Emission}

\noindent To simulate the detection process involving the Roman telescope for surveying microlensing events and subsequent follow-up observations with the SKA telescope, a model is necessary to estimate the emission of exoplanets based on their characteristics and the attributes of their parent stars. In general, the flux received from a planet contains thermal radiation due to its intrinsic temperature, as well as the reflection radiation from the parent star and radio emission if it has a magnetic field due to the electron cyclotron maser instability (ECMI) \citep{melrose1982electron, dulk1985radio}. The thermal and reflection radiation can be readily computed from the planet's characteristics and its parent star \citep{bagheri2019detection}. %The electron cyclotron maser instability (ECMI) is a phenomenon observed in plasmas, where electrons emit coherent electromagnetic radiation as they are confined by magnetic field lines along their trajectory, undergoing gyration and acceleration. 
The ECMI radiation is characterized by strong circular polarization and high anisotropy. ECMI stands out as the most efficient mechanism for radio emission, superseding other mechanisms like beam-plasma instabilities. Consequently, it plays a dominant role in the generation of radio emissions. This dominance is exemplified in the brightness of auroral radio emissions from planets in the solar system, particularly Jupiter, which rival the intensity of solar radio bursts.\\
\\
%The electron cyclotron maser instability is a phenomenon that occurs in plasmas where electrons emit coherent electromagnetic radiation due to confinement by magnetic field lines along which they travel, gyrate, and accelerate.  ECMI radio emission is strongly circularly polarized and highly anisotropic.  The ECMI is the most efficient radio emission mechanism, dominating all other emissions (beam-plasma instabilities and others).  This is why the auroral radio emissions of solar system planets – especially that of Jupiter – are as bright as Solar radio bursts.
Based on observations of the magnetized Solar system planets, we know that all auroral %{or satellite-Jupiter} 
radio emissions power is related to incident energy flux of the stellar wind \citep{zarka2001magnetically, zarka2004fast, zarka2007plasma, zarka2018jupiter}. In all cases in our solar system, about ~$2 \times 10^{-3}$ of the electrons' energy goes into radio waves \citep{zarka2018jupiter}. Therefore, the average ECMI radiation can be linked to both the total power of the solar wind and the magnetic field strength of the planet, utilizing the RBL model \citep{zarka2001magnetically, zarka2004fast, zarka2007plasma}.  The gyrofrequency can be expressed mathematically as:
\begin{equation}
\nu = \frac{eB}{2 \pi m}
\end{equation}
where $\nu$ is the emitted pick frequency, $e$ is the charge of the emitting particles (in this case, electrons), $B$ is the magnetic field strength, and $m$ is the mass of the emitting particles (electron mass in this case).  This equation shows that the emitted frequency is directly proportional to the magnetic field strength of the planet.  Therefore, observations of this emission type can provide an indirect means of measuring and classifying the planetary magnetic field.  This quantification of the magnetic field gives us constraints on internal structure models and planetary rotation, informing us about the evolutionary history of the planet and host system \citep{hess2011model}.
\\
\\
The RBL model is frequently used to estimate the radio emission of extrasolar planets \citep{farrell1999possibility, zarka2001magnetically, joseph2004radiometric, zarka2007plasma, christensen2010dynamo}. 
The simplicity of this model has some advantages, but it does not provide a complete picture of all processes involved, in particular for close-in exoplanets as discussed in \citep{bagheri2024fresh} (for Examples of studies that extend beyond the RBL model for calculating radio emission \citep{vidotto2017predicting, li2023statistical, dong2018atmospheric, vidotto2017stellar}). However, the RBL model facilitates comparative analyses between the radio emissions of exoplanets and those observed within the solar system, thereby enhancing our contextual understanding of planetary magnetospheric dynamics. This capacity for comparative analysis underscores the utility of the RBL model in advancing our comprehension of exoplanetary environments and their potential habitability while also serving as a foundation for future investigations into the broader astrophysical implications of planetary radio emissions. Thus, in this paper, we follow the approach outlined by \citep{joseph2004radiometric} to estimate the radio flux density of exoplanets. As described by \citep{farrell1999possibility}, the radio power from an exoplanet is related to the incident power of the stellar winds. Following \citep{farrell1999possibility}, the Jovian decametric component is considered at least partly related to the solar wind kinetic energy input and is used as the base power level to give
\begin{equation}
    P_{rad} = 4 \times 10^{18} erg~ s^{-1} \big(\frac{{\dot{M}_{ion}}}{10^{-14} M_\odot ~yr^{-1}}\big)^{0.8} \big(\frac{v_\infty}{400 ~km ~s^{-1}}\big)^{2} \big(\frac{a}{5 ~au}\big)^{-1.6} \big(\frac{\omega}{\omega_J}\big)^{0.8} \big(\frac{M_p}{M_J}\big)^{1.33}~,\label{eqPower}
\end{equation}
where $a$ is the semi-major axis of the planet’s orbit in au, $\dot{M}$ ion is the stellar ionized mass-loss rate, $v_\infty$ is the terminal velocity of the stellar wind, and $\omega$ is the corotation speed. For the terminal velocity of stellar wind, we use
\begin{equation}
     v_{\infty} = 0.75 \times 617.5 \times \sqrt{R_\star/M_\star}~, \label{eq:terminal}
\end{equation}
where $R_\star$ and $M_\star$ are the host star radius and mass in solar radius and mass units \citep{o2018search}. Equation \eqref{eq:terminal} serves as a mere approximation for low-mass stars, however, recent study \citep{mesquita2020global} indicates that the terminal velocity of M dwarf stellar winds either aligns well with predictions from Parker wind models, in which stellar winds are driven by radiation pressure mechanism, or exceed these model estimates. As a result, employing this equation may yield conservative estimations of stellar winds terminal velocity and the overall count of planets with detectable signals. Then, the radio flux density ($S_\nu$) emitted by an exoplanet can be estimated using the following equation
\begin{equation}
    S_\nu = \frac{P_{rad}}{\Delta \nu \Omega d^2}~, \label{eqFreq}
\end{equation}
where $P_{rad}$ is the total radio emission power of the exoplanet \eqref{eqPower}, $d$ is the distance between the exoplanet and the observer, and $\nu$ is the frequency of the radio emission,
 \begin{equation}
    \nu_c = 23.5 ~\text{MHz} ~\Big(\frac{\omega}{\omega_J}\Big) \Big(\frac{M_p}{M_J}\Big)^{5/3} \Big(\frac{R_p}{R_J}\Big)^{3}~, \label{eqPlanetfrequency}
\end{equation}
where $R_p$ is the planetary radius. 
Substituting \eqref{eqFreq} into \eqref{eqPower} gives 
\begin{eqnarray}
    S_\nu &=& 7.6~ \text{mJy}~ (\frac{\omega}{\omega_J})^{-0.2} (\frac{M_p}{M_J})^{-0.33} (\frac{R_p}{R_J})^{-3}
    (\frac{\Omega}{1.6 ~ sr})^{-1} (\frac{d}{10~pc})^{-2}
    \\
    &\times& (\frac{a}{1 ~au})^{-1.6}
    (\frac{{\dot{M}_{ion}}}{10^{-11}~M_\odot ~ yr^{-1}})^{0.8} (\frac{v_\infty}{100~ km~s^{-1}})^{2}~,
\end{eqnarray}
where $\omega_J$, $M_J$, and $R_J$, are Jupiter's corotation speed, mass, and radius and $\Omega$ is the beaming solid angle of the emission. In deriving this expression, we have followed \citep{farrell1999possibility} and assumed that the planet will emit ECMI emission between the frequencies $0.3 \nu_c$ and $\nu_c$, where $\nu_c$ is the maximum radiation frequency. \\
\\
The planetary corotational speed ($\omega$) is a crucial parameter in estimating radio flux from exoplanets. Exoplanets orbiting closely around their host stars experience significant tidal dissipation, resulting in considerably slowed corotation or complete tidal locking. This phenomenon has a profound impact on the production of auroral radiation due to the permanent orientation of one side of the planet towards the star, causing asymmetrical ionospheric conductance \citep{zarka2001magnetically, seager2002constraining}. To consider the effect of tidal locking, we assume all the exoplanets with orbits $< 0.1$ au are tidally locked, so the orbital period is equal to their corotation period. For all other exoplanets, we use the Darwin-Radau relation to relate corotation and oblateness \citep{murray2000solar} 
\begin{equation}
\omega = \sqrt{\frac{f~G~M_p}{R_{eq}^2}\Bigg[\frac{5}{2}(1-1.5~C)^2 + \frac{2}{5}\Bigg]}  
\end{equation}
where $C$ is 0.4 for rocky planets and for gas giant planets $C = 0.25$ \citep{hubbard1984planetary}. In this equation, $f$ is the planet's oblateness; the oblateness of planets refers to the degree to which a planet deviates from a perfect sphere, taking on a slightly flattened shape due to the centrifugal force generated by its corotation. This deviation from sphericity is more pronounced for rapidly rotating planets. The oblateness of a planet is quantified by 
\begin{equation}
    f = \frac{R_{eq} - R_{pole}}{R_{eq}}
\end{equation}
where $R_{eq}$ and $R_{pole}$ are the equatorial and polar radii of the planet. In our simulation, for the Jovian planet with orbital distance $a > 0.1$ au, we use the oblateness of Jupiter ($f = 0.064$), and for rocky planets, we use the Earth's value which is $f = 0.00335$ \citep{barnes2003measuring}. In Our simulation we also assume no planets rotate faster than Jupiter.
%For close-in exoplanet, this has a value between $10^{-5} - 10^{-3}$ \citep{}. In our simulation for the Jovian planet with orbital distance $0.1 < a < 0.5$ au we randomly choose a value between this range, for Jovian planets with orbits greater than 0.5 au, we use the oblateness of Jupiter ($f = 0.064$), and for rocky planets, we use the Earth's value which is $f = 0.00335$ \citep{barnes2003measuring}.
\\
\\
%The fundamental assumption in our simulations posits the presence of an exoplanet in orbit around the source star. As a consequence, the observed flux is a composite of the fluxes emanating from both the source star and its accompanying planet. \\
In this paper, the arrangement considered for the lens, source star, and planet entails that the observed light curve of microlensing events results from a combination of two distinct light curves: one originating from the source star and the other from the planet. Consequently, the final light curve results from merging these two light curves. To calculate host stars' radio flux at the desired frequency, we use the relationship between the luminosity of the host star in the X band and the radio band.  Stars emit X-rays primarily through high-energy processes such as magnetic reconnection events, coronal heating, and interactions between fast-moving charged particles and the stellar atmosphere. The X-ray luminosity of a star is thus influenced by its magnetic activity level, mass, age, and evolutionary stage.
On the other hand, radio emission from stars is more associated with non-thermal processes, such as ECMI. The strength of radio emission depends on factors including the density and strength of the stellar magnetic field, the presence of energetic particles, and the efficiency of acceleration mechanisms. While there is no universal scaling relationship between X-ray and radio luminosity for all types of stars, correlations have been observed within specific stellar populations, such as active stars with strong magnetic fields or those undergoing rapid evolution. One can use empirical relationships based on observations for the radio luminosity of main sequence stars. One such relationship is the Gudel-Benz relation, which relates the X-ray and radio luminosities:
\begin{equation}
L_X = L_R \times 10^{15 \pm 1}
\end{equation}
where $L_X$	is the X-ray luminosity and $L_R$ is the radio luminosity \citep{guedel1993x}. Notably, ECMI radiation exhibits strong circular polarization, whereas solar or stellar plasma radiation lacks polarization and occurs sporadically. Additionally, ECMI radiation is beamed anisotropically, resulting in significant modulation by the planetary rotation \citep{zarka2014magnetospheric}. Therefore, polarization and temporal variations should enable the differentiation between stellar and exoplanetary radio emissions, as gravitational lensing does not change the polarization of the light rays. \\

\subsection*{Induced Radio Emission by Exomoons}
\noindent There are more than 200 moons in our solar system, many of which orbit gas giants like Jupiter and Saturn. Despite significant efforts, the successful detection of exomoons remains elusive \citep{narang2023ugmrt}. If an exomoon resides within the magnetosphere of its host planet, it can influence the planet's radio emissions by injecting plasma that accelerates electrons within the magnetosphere. A notable example is Io, a moon characterized by intense volcanic activity within Jupiter's magnetosphere. This activity generates a tenuous atmosphere around Io \citep{lopes2007summary}, contributing ions to Jupiter's magnetosphere and generating a plasma \citep{reich2009cosmic}. These ions in motion accelerate electrons, ultimately leading to the formation of an ECMI along the field lines, resulting in radio emissions known as Io-DAM (see Figure 1-a in \citep{cecconi2014goniopolarimetry}) \citep{crary1997generation, mauk2001understanding}. Similar effects are observed with other Galilean moons and moons orbiting Saturn, such as Titan and Enceladus. Consequently, it is reasonable to anticipate moon-planet interactions in other stellar systems \citep{green2021magnetospheres}. ECMI radio emissions from exomoon magnetosphere have not had a confirmed detection, but there are three candidate systems under study \citep{turner2021search}.\\

\noindent To simulate the induced radio emission by exomoons, one should know the probability of having a moon such as Io. The results of N-body simulations in \citep{heller2014formation, sasaki2010origin} show that 80\% of Jovian planets have moons with masses $M_m/M_p ~\approx 10^{-5}$. However, stellar tidal forces serve to slow down a planet's rotation and in conjunction with tidal migration, contribute to removing moons; At short $a$, migration of moons is significant and occurs quickly, which can lead to orbital instabilities \citep{sucerquia2020can}. This process limits the lifetimes of larger moons in extrasolar planetary systems \citep{barnes2002stability}. 
Therefore, in our simulation, we assume 80\% of Jovian planets with masses $M_p \geq 0.1~M_J$  with $a \geq 0.1 \text{au}$ have one moon. So exoplanets with the semi-major axis of less than 0.1 au do not have any moons, which is consistent with assuming they are tidally-locked. \\
\\
\citet{noyola2014detection} proposed that interactions akin to those between Io and Jupiter's magnetosphere could also occur in exoplanet-moon pairs. The frequency of the moon-induced radio emission is slightly different,
\begin{equation}
    \nu_{c_i} =\nu_c \times \sqrt{4 - 3 \frac{R_p}{a_m}}~, \label{eq:moonfrequency}
\end{equation}
where $\nu_{c_i}$ is the frequency of radio emission induced by a moon, such as Io, $a_m$ is the same-major axis of the moon's orbit, and $\nu_c$ can be obtained by \eqref{eqPlanetfrequency}. Equation \eqref{eq:moonfrequency} is asymptotic, approaching a maximum value of $\nu_c$ as $a_m$ becomes very large \citep{noyola2016radio}. Assuming the plasma density of the exomoon is equal to the Io's plasma density, the optimal distance for the exomoon to have the maximum of the induced radio emission can be obtained by \citep{noyola2014detection}
\begin{equation}
    a_{{m}_{optimal}} = 5.4 \times M_p^{0.3}~.
\end{equation}
\noindent Therefore, the total radio flux induced by a moon is
\begin{equation}
    S_{\nu_i}= \frac{2\pi \beta R_m^2 B_m^2 V_0}{\mu_0 ~\nu_{c_i} ~\Omega ~d^2} \sqrt{\frac{\rho_m}{\rho_m + \rho_c}}~ ,
\end{equation}
where $d$ is the distance of star-planet-moon system to Earth, $\rho_m$ is the moon's plasma density, $R_m$ is the moon's radius, and $\beta$ is the efficiency coefficient of converting Joule dissipation to radio waves (we assume $\beta \approx 1\%$ as is in the Jupiter–Io system \citep{zarka2001magnetically}). $B_m$ is the magnetic field affecting the exomoon, and so $B_m = B_{p} \times (R_p/a_m)^3$, where $B_{p}$ is the magnetic field of the exoplanet at its poles. The plasma speed, $V_0$, is computed assuming it corotates with the planet’s magnetic field. Therefore, $V_0 = \omega_p \times a_m - \sqrt{G~M_p/a_m}$, where $G$ is the gravitational constant, $M_p $is the planet’s mass, and $\omega_p$ is the planet’s corotation speed. Finally, $\rho_c$ is the critical density which is defined as $\rho_c = \mu_0^{-1}(B_m/V_0)^2$ \citep{noyola2014detection}.  In our simulation, we assume IO's plasma density and IO's radius for $\rho_m$ and $R_m$. We also assume $\Omega = 0.2$ \citep{noyola2014detection}. In our simulation, the induced radio emission by an Io-like moon accounts for a maximum of 0.2\% of the total radiation emitted by the exoplanets.

\subsection{Observation Strategy}

\noindent The NASA Roman Space Telescope is a next-generation space observatory designed to investigate various astrophysical phenomena using infrared observations. Named in honor of the astronomer Nancy Grace Roman, often regarded as the "Mother of Hubble" for her pivotal role in the development of the Hubble Space Telescope \citep{roman2019nancy}, this advanced instrument is slated to launch in the mid-2020s. With its wide field of view, the telescope is designed to conduct large-scale surveys to explore dark energy, dark matter, and the formation and evolution of galaxies with unprecedented detail \citep{rose2021reference}. The Roman Space Telescope will also be crucial in advancing our knowledge of exoplanets using microlensing events. Roman W149 filter ($0.9-2~\mu m$) will have a significantly low detection threshold, with a zero-point magnitude of $27.61$. The microlensing survey of Roman will monitor $1.97 \text{deg}^2$ of the Galactic bulge (in the direction of $b = -1.5$ and $l = 0.5$ in the Galactic coordinate) with 15-minute cadence, over six 72-day per season \citep{spiegel2005can}, potentially detecting thousands of exoplanets via the perturbations that they produce on the microlensing light curves \citep{bagheri2019detection}. Since microlensing events usually have a duration of a few days up to a few weeks, we propose to use the Roman telescope as a survey observer to detect a microlensing event and then alert that event to a ground-based radio telescope for a follow-up observation of that microlensing event in the radio band. The most suitable ground-based telescope to detect planetary radio emission is 
the Square Kilometre Array (SKA) telescope. As the world's largest radio telescope, the SKA comprises two arrays: SKA1-Low, optimized for low-frequency observations between 50 MHz and 350 MHz, and SKA1-Mid, designed for mid-frequency observations between 350 MHz and 14 GHz. The vast collecting area of the SKA, totaling one square kilometer when fully operational, enables it to detect faint radio signals from cosmic phenomena spanning a wide range of scales, from nearby planets to distant galaxies and beyond. Furthermore, the SKA's innovative design incorporates advanced signal processing techniques and data analysis algorithms, maximizing its scientific output while minimizing data volumes and processing requirements. 

%As a collaborative endeavor involving contributions from over 20 countries, the SKA embodies the spirit of international cooperation and scientific excellence, poised to unlock new frontiers in radio astronomy. 

\subsection{Detectability of the Planetary Radio Signal}

\noindent The detectability of the radio signal of exoplanets depends on, first, the detection of the microlensing event by the Roman telescope, and, second, the frequency and the intensity of the radio emission of the star-planet system. To address the first condition, we calculate the host star and planet magnitude in Roman's W149 filter. For the flux received from the host stars, we consider both the distance-dependent effects of the source stars and the reddening of their apparent magnitudes due to interstellar dust. Using the host stars' position and stellar type, we estimate the extinction caused by interstellar dust along the line of sight, using the comprehensive 3D extinction map provided by \citep{marshall2006modelling}. We use the relationships outlined in \citep{nishiyama2008interstellar, nishiyama2009interstellar} to convert the extinction values from V bands to the band of W149 filter \citep{bagheri2019detection, penny2019predictions}. The flux received from a planet comprises both thermal radiation originating from the planet's intrinsic temperature and reflected radiation from its parent star. The planet's temperature can be calculated by assuming that the planet's thermal emission adheres to blackbody radiation principles \citep{lopez2007thermal}. This calculation incorporates the absorption of radiation from the parent star by the planet, which is then re-emitted following Boltzmann's law,
\begin{equation}
    T_p = T_{eff} \sqrt{\frac{R_\star}{a}}~\Big[f \times (1-A_B)\Big]^{1/4}
\end{equation}
where $T_p$ indicates the temperature of the planet, $A_B$ denotes the albedo (0.15 and 0.52 for rocky and Jovian planets), while $f$ characterizes the proportion of re-radiated energy absorbed by the planet \citep{harrington2006phase, knutson2007map}.  Since hot Jupiters are the most frequently detected, we adopt a value of $2/3$ for $f$ across all planets. Assuming blackbody radiation, the thermal flux ($F_{th}$) at the specified frequency $\nu$ can be expressed using Planck's law as follows:
\begin{equation}
I(\nu, T_p)= \frac{2h~\nu^3}{c^2} \frac{1}{\exp(h\nu/k~T_p)-1}
\end{equation}
where, $I(\nu, T_p)$ denotes the emitted power per unit area of the emitting surface, per unit solid angle, and per unit frequency. Integrating it across the energy range of the W149 filter and over a hemisphere yields the thermal flux of the planet as observed by Roman. The flux of the planet resulting from the reflection ($F_{ref}$) of the parent star's light is determined by
\begin{equation}
F_{ref}= A_g~ g(\Phi) ~F_\star \Big(\frac{R_p}{a}\Big)
\end{equation}
where $g(\Phi)$ denotes the fraction of the illuminated portion of the planet visible to the observer, and $A_g$ represents the geometric albedo. We utilize a geometric albedo formula: $A_g = 2/3 ~A_B$ \citep{bagheri2019detection}. While this relationship may not hold true for all phase angles, its alteration does not significantly impact results. This is because the thermal radiation emitted by planets is much weaker than the flux of host stars in the Roman W149 filter. Thus, any changes in this relationship would only affect the total flux received by the Roman telescope at the order of $10^-4$ at most.
Consequently, the overall flux received from the planet ($F_p$) can be calculated as the combined sum of the thermal flux and the reflection flux. In the final stage, we add realistic noise. The total noise consists of the intrinsic Poisson fluctuation in the flux received from the microlensing event in a single exposure, the intrinsic Poisson fluctuation within the point spread function (PSF) representing the background sky flux in a single exposure, the read-out and dark noise \citep{bagheri2019detection}. Finally, we determine the range within which the noisy signal exhibits magnitudes below the zero-point magnitude of Roman and above the saturation magnitude of 14.8 \citep{penny2019predictions}. \\
\\
For the follow-up observation by the SKA, we consider the SKA1-low and SKA1-mid frequency ranges up to 890 MHz. The image noise is given by,
\begin{equation}
  \sigma = S_D~ \frac{SEFD}{\eta_S \sqrt{\eta_{pol}~ \Delta \nu \Delta \tau}},  
\end{equation}
where $S_D = 2.5$ is a degradation factor relative to the natural array sensitivity for the specific target Gaussian FWHM resolution of the image, $\eta_S = 0.9$ is a system efficiency that takes account of the finite correlator efficiency, and $\eta_{pol} = 2$ is the number of contributing polarizations and for SKA1 with orthogonal linear polarizations \citep{braun2019anticipated}. The imaging sensitivity for SKA1-Low and SKA1-mid up to 890 MHz is listed in Table \ref{tab:SKA} by assuming a continuum observation with fractional bandwidth of $\Delta \nu/\nu_c \approx 0.3$, together with an integration time $\Delta \tau = 1$ hour. 

\begin{longtable}{cccc}
\hline\hline
$\nu_{min}$ [MHz]   & $\nu_{mid}$ [MHz]   &$\nu_{max}$ [MHz]    &$\sigma$ [$\mu Jy$]
\\
\hline
50      &60         &69         &163\\
69      &82         &96         &47 \\
96      &114        &132        &26 \\
132     &158        &183        &18 \\
183     &218        &253        &14 \\ 
253     &302        &350        &11 \\
350     &410        &480        &16.8 \\
480     &560        &650        &8.1 \\
650     &770        &890        &4.4 \\
\hline
\caption{Image sensitivity of SKA1-Low and SKA1-mid within the indicated frequency bands for continuum observations \citep{braun2019anticipated}} \label{tab:SKA}
\end{longtable}
\noindent We use the SKA image sensitivity for different ranges of frequencies as summarized in  Table \ref{tab:SKA}. Therefore, any event with flux density greater than the SKA sensitivity for at least 4 hours of imaging is considered a detected event. 
\section{Results}
\begin{figure}[ht!]
\centering
\includegraphics[width=55mm]{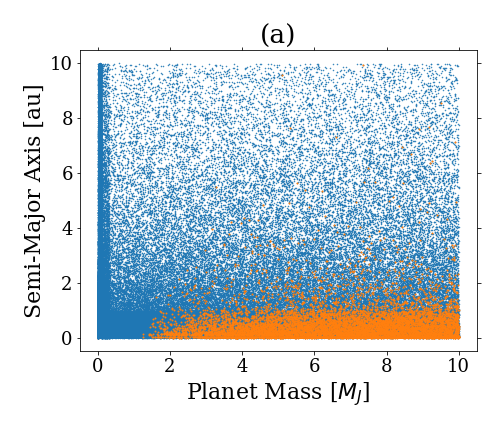}
\includegraphics[width=55mm]{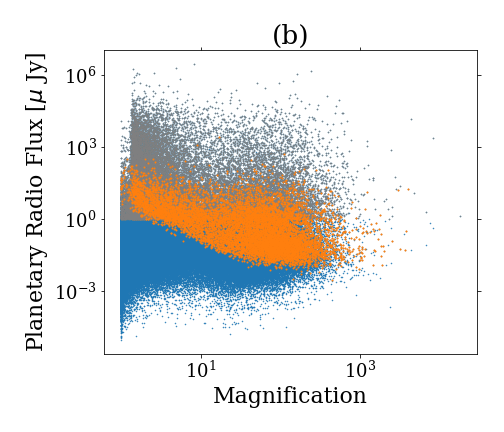}
\includegraphics[width=55mm]{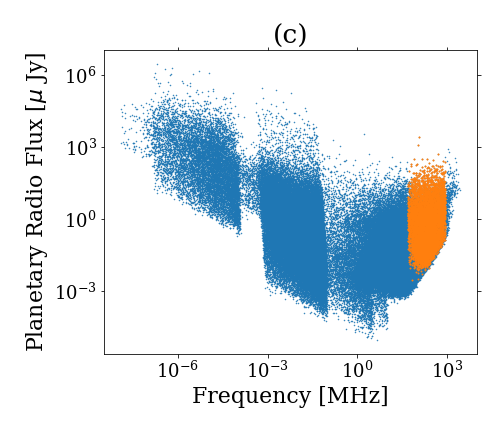}
\\
\includegraphics[width=55mm]{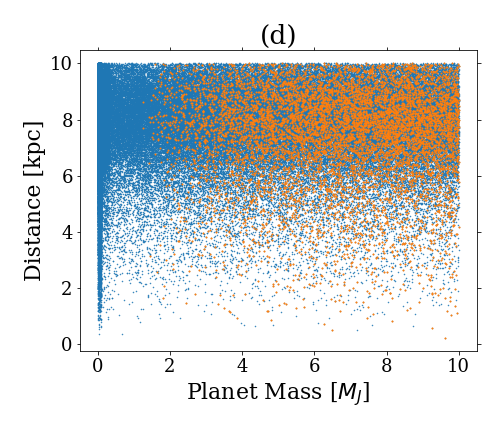}
\includegraphics[width=55mm]{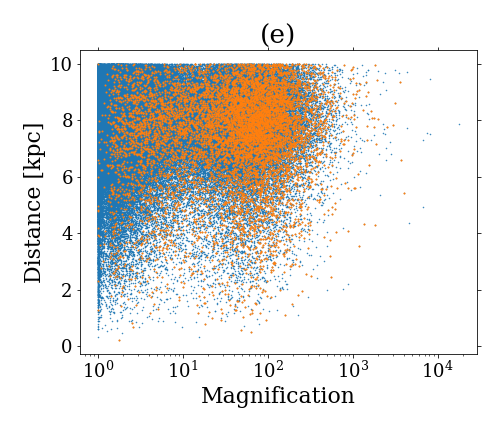}
\caption{Characteristics of the exoplanets within the simulated event. Exoplanets exhibiting detectable radio signals are highlighted in orange;
a) distribution of semi-major axis as a function of planetary mass, b) planetary radio flux vs. magnification of the microlensing events, c) planetary radio flux as a function of the frequency of the planetary radio emissions, d) distance of the star-planet system vs. planetary mass, and e) distance of the star-planet system vs. magnification of the microlensing events. \label{fig:scatterPlots}}
\end{figure}

\begin{figure}[ht!]
\centering
\includegraphics[width=70mm]{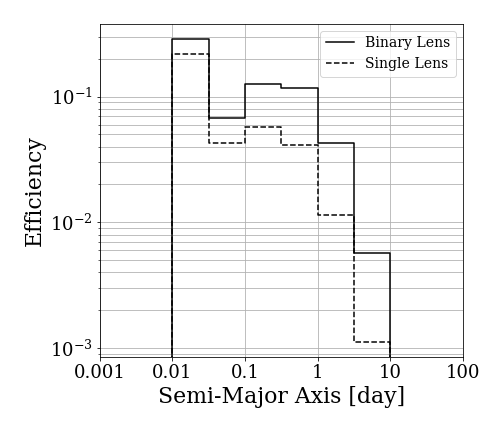}
\includegraphics[width=70mm]{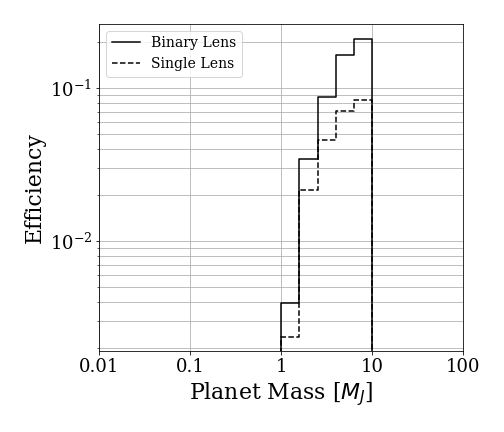}
\caption{The detection efficiency of the planetary radio emission in terms of the mass and semi-major axis of the exoplanet in our Monte Carlo simulation.
\label{fig:histoPlots}}
\end{figure}

We simulate 150,000 single-lens and binary-lens microlensing events. %Figure \ref{fig:eventSample} represents two of those that Roman can detect. 
Our Monte Carlo simulations indicate the likelihood of detecting radio emission of rocky planets using either single-lens or binary-lens configurations is exceedingly low with the SKA telescope. Consequently, Jovian planets exhibit the highest detection probabilities in both scenarios. This outcome primarily stems from their stronger magnetic field and higher frequency of radio emission relative to rocky planets, resulting in a substantial portion of their radiation falling within the SKA sensitivity. Simulations of Roman-SKA data predict, under the assumption of single-lens configuration, that the number of detecting a Jovian planet with orbital periods of 2 days to 10 years is about 3.54 \%. For the binary-lens simulation, this number increases to 8.25 \% due to the high magnification of caustic crossings. \\
\\
Figure \ref{fig:scatterPlots} shows our simulation's distributions of planetary parameters. Exoplanets with detected radio fluxes are in orange. As depicted in Figure \ref{fig:scatterPlots}-a, it's notable that the majority of exoplanets exhibiting detectable signals are situated at orbital distances lower than 2 au. However, for planets with higher masses, the orbital distance can extend up to 8 au.
%This decreasing abundance of the simulated high-mass planets with periods $< 1500$ days (corresponds to $a < 2.5$ au), as seen in Figure \ref{fig:scatterPlots}-a, results from the relationship between giant planets' abundance and the host stars' metallicity as outlined by \citep{johnson2010giant}. This relationship was established for planets with $a < 2.5$ au. These criteria typically lead to conservative estimates of the overall count of detected giant planets with smaller orbital periods. 
The planetary radio flux in terms of magnification of the microlensing events is shown in Figure \ref{fig:scatterPlots}-b. Exoplanets within the shaded region exhibit emission frequencies below 50 MHz, rendering their signals undetectable despite their radio emissions being of high intensity and possessing a significant magnification factor during microlensing events. Figures \ref{fig:scatterPlots}-d and \ref{fig:scatterPlots}-e represent the distance of the star-planet system relative to the exoplanet mass and the magnification of the microlensing events. Notably, the range extends up to 10 kpc for exoplanets with detectable signals, underscoring the importance of this observation method. These exoplanets are beyond the reach of current technologies for detection in any other wavelength bands using conventional methods. Identifying such star-planet systems can enhance our understanding of planet formation rates within the galactic bulge and facilitate the detection of extragalactic exoplanets.\\
\\
The detection efficiency in terms of mass and semi-major axis of the exoplanets is illustrated in Figure \ref{fig:histoPlots}. Detected planets exhibit masses ranging from $M_p = 1$ to $10 M_J$ and semi-major axes spanning from $a = 0.01$ to $10$ au.\\
\\
\noindent We conducted simulations targeting coordinates $l = 0.5^\circ$ and $b = -1.5^\circ$. In this direction, as reported by \citep{penny2019predictions}, the estimated number of source stars detectable by the Roman telescope with magnitudes below 25 is $240 \times 10^6$. Hence, the number of microlensing events that could be detected via Roman observation can be given by:
\begin{equation}
N = \frac{2}{\pi} T_{obs} \times N_\star \times < \epsilon(t_E) > \tau 
\end{equation}
where $T_{obs}$ is the observational time, $N_\star$ is the number of background stars detectable by Roman, $\epsilon(t_E)$ is the Roman efficiency for detecting microlensing events with the duration of $t_E$, and $\tau$ is the optical depth \citep{bagheri2019detection}. The optical depth, averaged over the distances of the source stars, is $\tau = 2.4 \times 10^{-6}$ \citep{mroz2019microlensing}, and our simulation yields an average value of $<\epsilon/t_E > = 0.035$. Therefore, around 27000 microlensing events can be detected by Roman observation \citep{penny2019predictions}. Consequently, approximately 1317 exoplanets could be detected with Roman-SKA observations toward the Galactic bulge. However, it may not be feasible to follow up on all Roman's detected microlensing events; nonetheless, a 10-20\% follow-up of these events could significantly impact exoplanet sciences and the measurement of exoplanetary magnetic fields. These findings are summarized in Table \ref{tab:resultSummary}.

%We repeat our simulation without considering the induced radio emission by exomoons. Figure \ref{fig:noMoons} represents the histograms of detected exoplanet parameters. As seen, the efficiency is much less than that of exomoons. 
\begin{center}
\begin{tabular}{ll}
\hline\hline
Estimated number of detectable planetary radio\\
emissions in the Roman-SKA follow-up observation\\
\hline
Stars (W149 $<$ 25)   &$240 \times 10^6$\\
Optical Depth   & $2.4 \times 10^{−6}$\\
Microlensing events &$27000$\\
Planet detection    &$\sim 1317$
%M_p$       & 0.002- 0.3 $M_J$    &0.3-1 $M_J$    &1-15 $M_J$  \\ 
%\#          & 0.002- 0.3 $M_J$    &0.3-1 $M_J$    &1-15 $M_J$  \\ 
%$a$         & 0.05–0.1 au         &0.1–0.5 au     &0.5–1 au    \\ 
%\#          & 0.002- 0.3 $M_J$    &0.3-1 $M_J$    &1-15 $M_J$  \\ 
%$\nu_c$     & 0.002- 0.3 $M_J$    &0.3-1 $M_J$    &1-15 $M_J$  \\ 
%\#          & 0.002- 0.3 $M_J$    &0.3-1 $M_J$    &1-15 $M_J$  \\ 
%$S_\nu$     & 0.002- 0.3 $M_J$    &0.3-1 $M_J$    &1-15 $M_J$  \\ 
%\#          & 0.002- 0.3 $M_J$    &0.3-1 $M_J$    &1-15 $M_J$  \\ 
%\hline
%\caption{The predicted numbers of detectable planetary radio signals in different ranges of the mass ($M_p$), semi-major axis ($a$), maximum of frequency ($\nu_c$), and radio flux ($S_\nu$) based on our Monte
%Carlo simulations of Roman-SKA observation. } 
\label{tab:resultSummary}
\end{tabular}
\end{center}

\section{Discussion}

\noindent Our comprehension of planetary magnetic fields remains largely confined to the solar system. Expanding our knowledge and observation of exoplanetary magnetic fields could provide valuable insights into their internal compositions and dynamics, which are otherwise challenging to ascertain. The detection of radio emissions from confirmed exoplanets has been investigated in multiple studies (e.g., \citet{2024arXiv240414468B, li2023statistical}. In this study, we investigate an alternative approach to detecting exoplanets through their radio emissions magnified by microlensing events. Our findings reveal that the planetary radio emissions of approximately 5\% of microlensing events could be detected through follow-up observations with Roman-SKA (equivalent to $\sim 8\%$ for binary-lens and $\sim 3.5\%$ for single-lens events). This finding gains significance when considering that, to date, only 210 exoplanets have been detected via microlensing events\footnote{NASA Exoplanet Archive}. These detections primarily occurred due to caustic crossings in binary-lens events, making follow-up observations of these planets nearly impracticable. In contrast, we consider exoplanets as companions within the source system of microlensing events, enabling us to conduct follow-up observations on the star-planet system.\\ 
\\
Our results underscore the potential of infrared-radio follow-up observations to enhance the rates of direct exoplanet detections while yielding valuable insights into the magnetic fields and internal structure of exoplanets. The observation strategy outlined in this study demonstrates heightened sensitivity to detecting the radio emissions of giant and close-in exoplanets. Observing the radio emissions of close-in exoplanets broadens our comprehension of planetary responses to their space environment and evolutionary trajectories. Hot Jupiters' magnetospheres serve as compelling subjects for study, highlighting the intricate complex interplay between magnetic fields, stellar wind, and atmospheric escape processes. Observing radio emissions of close-in exoplanets and further investigations into these magnetospheric properties will deepen our understanding of exoplanetary systems and their diverse environments. Such insights will significantly contribute to our comprehension of planet formation, migration, and the potential habitability of exoplanets.\\
\\
The outcomes of this project are important for the prospects of Exoplanet radio emission detection by the contemporary and planned surveys as well as the next generation of Radio telescopes. In a wider view, the results of this work pave the way toward actual infrared-radio follow-up observations of exoplanets/exomoons using the method introduced in this study. The knowledge of the magnetic field and the magnetospheric emissions of exoplanets are considered a science frontier for the next decade, as highlighted in the Origins, Worlds, and Life Planetary Science \& Astrobiology Decadal Survey report. Two Priority Science Question Topics identified in this survey report incorporate aspects of planetary magnetic fields and their interplay with the solar wind. These are identified as Q6: "Solid body atmospheres, exospheres, magnetospheres, and climate evolution" and Q12.7: "Exoplanets, Giant planet structure and evolution."\citep{national2022origins}. 
\\
\\
%This will open a new window to exoplanetary research.
%The proposed research plan can pave the way toward the first successful detection of an exomoon.  \\
%\\
%Examples include Jovian planets around the young stars [Wood et al., 2002, 2005] or even indirect detection of exomoons via their effects on their parent exoplanets
\section*{acknowledgments}
\noindent We acknowledge the use of the Cheyenne supercomputer at the National Center for Atmospheric Research (NCAR) \citep{computational2017cheyenne} and Texas Advanced Computing Center (TACC) supercomputing resources. We acknowledge the support of the US National Science Foundation (NSF) under Grant No. 2138122.

\bibliographystyle{aasjournal} %  Many Frontiers journals use the Harvard referencing system (Author-date), to find the style and resources for the journal you are submitting to: https://zendesk.frontiersin.org/hc/en-us/articles/360017860337-Frontiers-Reference-Styles-by-Journal. For Humanities and Social Sciences articles please include page numbers in the in-text citations
\bibliography{main}

\end{document}